\documentstyle[twoside,fleqn,espcrc2]{article}

\newcommand{\AmS}{{\protect\the\textfont2
  A\kern-.1667em\lower.5ex\hbox{M}\kern-.125emS}}

\title{ Magnetic Monopoles in non-compact QED -
 is there a Phase Transition? }

\author{P.E.L. Rakow \address{Institut f\"ur Theoretische Physik,
        Freie Universit\"at Berlin,\\
        Arnimallee 14, 1000 Berlin 33, Germany}}

\begin{document}

\begin{abstract}
  The existence of the monopole condensation transition
 reported by Koci\'c et al.~\cite{newquench} in
 non-compact, quenched QED is tested. No phase transition is found.
 This shows that divergence of the `monopole susceptibility'
 introduced by Hands and Wensley \cite{Hands} is not a reliable
 indicator of second order phase transitions. In view of these results
 I discuss claims \cite{confine} that the chiral phase transition
 seen in QED with fermions \cite{confine,us} is a lattice artefact
 driven by monopole condensation.
\end{abstract}

\maketitle

\section{Introduction}

    The authors of \cite{newquench,Hands} have
 investigated magnetic monopoles in non-compact, quenched QED and find an
 `authentic second order phase transition' at $\beta_c \approx 0.244$
 (see for example ``The Universality Class of Monopole
 Condensation in Non-Compact, Quenched QED" \cite{newquench}).
 This separates a weak coupling phase, in which monopoles are rare
 and unimportant, from a strong coupling phase in which the monopoles
 condense. The evidence for this phase transition comes from a monopole
 cluster susceptibility defined in \cite{Hands}. This is reminiscent
 of the behaviour seen in compact QED \cite{DGT}, where the
 strong coupling phase is a dual superconductor which confines
 electrical charge. (Dual superconductivity and  charge
 confinement are to be expected whenever monopoles condense.)

    The occurrence or non-occurrence of a second order monopole phase
 transition is important because such a transition would
 imply the existence of monopoles in the theory's continuum limit.
 This would cast doubt on conclusions about continuum physics drawn
 from lattice calculations, as monopoles are (presumably) absent
 in reality.

    I test the scenario of \cite{newquench,Hands} with
 a number of analytic results. I show that
 quantities such as the density of monopoles and of Dirac string show
 no singularity at any finite $\beta$. I also look at an order parameter
 for monopole condensation (the conventional monopole susceptibility)
 which shows that the monopoles never condense.

  The monopole cluster susceptibility is closely related to a
 quantity widely used in studies of percolation \cite{percol}.
 It shows where the percolation of monopole
 world lines first occurs. However in the final
 section I argue that this monopole percolation
 threshold can have no effect on any physical quantities, and
 so can not be the cause of a phase transition.
 \raisebox{8.75cm}[0cm][0cm]{Preprint FUB-HEP 22/92}
 \raisebox{8.6cm}[0cm][0cm]{\hspace{34mm} December 1992}\vspace{-5mm}

\section{Analytic Results \label{analytic}}

\subsection{Defining monopoles on the lattice}

   The action for non-compact QED is
\begin{equation}
  S = \frac{1}{2} \beta \sum_{\mbox{plaqs}} F_{\mu\nu}^2
\end{equation}
  where $F_{\mu\nu}$ is the usual non-compact plaquette.
 $F_{\mu\nu}$ is linear in the $A_\mu$ fields so this action is Gaussian.

   Because charged fields
 couple only to the compact part of the gauge field there is some
 motivation for looking at monopoles. To define monopoles \cite{DGT,Hands}
 the plaquette $F_{\mu\nu}$ is decomposed into an integer valued `string'
 field $N_{\mu\nu}$ and a compact field $f_{\mu\nu}$ which lies in the
 range $(-\pi,\pi]$.
 \begin{equation}
   F_{\mu\nu}  = 2\pi N_{\mu\nu}  + f_{\mu\nu}
 \end{equation}

  The Bianchi identity tells us that $F$ summed over
 any closed surface always gives zero. This doesn't apply to the
 $N$ and $f$ fields separately. If the six faces of a cube are summed
 over
\begin{eqnarray}
   \sum_{\mbox{faces}} N_{\mu\nu}  =
  -\frac{1}{2\pi } \sum_{\mbox{faces}} f_{\mu\nu} \equiv M \label{mrange}
 \\[2mm]
  M \in \{ -2,-1,0,1,2\}  \nonumber
 \end{eqnarray}
   The integer $M$ is called the monopole charge.
  \raisebox{-2cm}[0cm][0cm]{\hspace{-7.5cm}
 \small{ Talk given at {\it Lattice 92},
 International Symposium on Lattice Field Theory, Amsterdam, 1992}}

\subsection{Monopole density  \label{dens_sect} }

    Because the action is Gaussian we can derive analytic formulae for
 most quantities.
 For example to calculate a formula for the monopole density I start
 with the full field distribution for the lattice, which is known from
 action. I then pick a particular cube and integrate out
 all fields that are not in this particular cube.
 Finally I integrate out the pure gauge modes.
 This leaves a probability
 distribution $\Psi$ for the six $F$ fields on the cube's surface.
 (In the following formulae I label the  outwardly directed $F$
 fields as the faces of a dice are labelled, i.e.~with the convention
 that $F_n$ and $F_{7-n}$ are on opposite faces).
 \begin{eqnarray}
 \lefteqn{\Psi(F_1,F_2,F_3,F_4,F_5,F_6) =} \nonumber \\
 &   c \, \beta^{5/2} \, \delta(F_1+F_2+F_3+F_4+F_5+F_6) & \nonumber \\
 &\times \exp \left[ -\beta x (F_1^2+ \cdots +F_6^2)
                                      \right. \hspace*{8mm}& \nonumber \\
 &  \hspace*{1.2cm}\left. -2 \beta y (F_1 F_6 + F_2 F_5 + F_3 F_4)\right]&
 \label{psi} \end{eqnarray}
  where $x$ and $y$ (which are the results of four-dimensional
  integrals) are
 \begin{eqnarray}
  x = 0.826\,049\,306\,\cdots,&y = -0.052\,878\,794\,\cdots
 \end{eqnarray}
 Note that the probability distribution Eq.(\ref{psi}) is of course
 a Gaussian, and that it obeys the Bianchi identity, as can be seen
 from the $\delta$ function.
 This distribution gives us all quantities that only involve
 one cube. The monopole density  $\rho(\beta)$ is
 \begin{eqnarray}
  \lefteqn{\rho(\beta) \equiv \langle \left|M\right| \rangle =}
                                                  \nonumber \\[2mm]
 \lefteqn{\int_{-\infty}^{\infty} \! \! \! dF_1 \cdots dF_6 \,
  \Psi\left(F_1, \ldots, F_6\right) \,
  \left|N_1+ \cdots +N_6\right|}            \label{monodens}
 \end{eqnarray}

   A similar formula gives the string density
 $\sigma(\beta)$. In Fig.\ref{fig:mono}
 these formulae are checked against the Monte-Carlo data of \cite{Hands}.
 ($\bullet$ shows $\rho(\beta)$ and $\scriptstyle{\triangle}$ shows
 $\sigma(\beta)/\rho(\beta)$. The error bars are smaller than the symbols.)
 Additional points were measured by
 G. Schierholz (personal communication). The agreement is excellent.
\begin{figure}[htb]
\vskip 4.7 truecm
\noindent \special{fig1.ps}
\caption{Monopole and string densities.}
\label{fig:mono}
\end{figure}

   It is of course much easier and more accurate to evaluate the
 five-dimensional integral Eq.(\ref{monodens}) than to evaluate a
 500\,000-dimensional integral by Monte-Carlo, but a more important
 advantage of the formula is that it allows us to check for the
 singularities that must occur at an `authentic second order phase
 transition'.

   The fact that $\left|M\right|$ is bounded Eq.(\ref{mrange}) is enough
 to show that all derivatives of the monopole density $\rho(\beta)$
 are finite at all $\beta$ values. If $\rho(\beta)$ is expanded as a
 series of the form
 \begin{eqnarray}
  \rho(\beta) =
     \beta^{5/2} \sum_{n=0}^{\infty} a_n \left(\beta_0-\beta\right)^n
 \label{series}\end{eqnarray}
 about an arbitrary point $\beta_0$ then the bound on $\left|M\right|$
 leads to bounds on the $a_n$.
 \begin{eqnarray}
   0 < a_n <
   \frac{1}{3}\,\frac{(2n+3)!}{n!(n+1)!}\frac{1}{4^n} \beta_0^{-5/2-n}
 \end{eqnarray}
 These bounds are strong enough to show that the series Eq.(\ref{series})
 is convergent with a radius of convergence of (at least) $\beta_0$.
 A convergent series expansion rules out the existence of any
 essential singularities in $\rho$.
 There is certainly no sign of a phase transition in $\rho(\beta)$.

\subsection{Monopole Susceptibility \label{susc_sect} }

   The standard definition of monopole susceptibility is
 \begin{eqnarray}
 \lefteqn{\chi_s = \big( \,\langle \sum_{x,y,\nu} M^\nu(x) M^\nu(y)\rangle }
                                        \nonumber \\
  &\hspace*{1cm}\left. - \sum_\nu
           \,\langle\sum_x M^\nu(x) \rangle^2 \right)/N_l&
 \label{standard}
 \end{eqnarray}
  where $M^\nu$ is the monopole current and $N_l$ the number of
 dual links in the lattice \cite{DGT}. This definition is
 completely analogous to other susceptibilities (such as the
 magnetic susceptibility of the Ising model or the topological
 susceptibility of SU(N)), and has a clear connection to the
 onset of Bose-Einstein condensation. When condensation takes
 place it can be seen in the fact that the fluctuations of
 boson number in a region of volume $V$ grow faster than the
 $V^{1/2}$ behaviour seen in the normal phase. An equivalent
 sign of condensation is that the boson density becomes infinitely
 sensitive to changes in chemical potential. The susceptibility
 Eq.(\ref{standard}) diverges if this happens, and so is an
  order parameter for condensation. Because $\chi_s$
  is a zero momentum Greens function it is clear that it can only
  diverge if the correlation length diverges, so infinite $\chi_s$
 implies the existence of a continuum limit. On the other hand
 the cluster susceptibility of \cite{Hands} does not have the
 form of a Greens function and has no obvious connection with
 Bose-Einstein condensation.

\begin{figure}[htb]
\vskip 4 truecm
\noindent \special{fig2.ps}
\caption{The monopole correlation function in momentum space.}
\label{fig:chi}
\end{figure}
  The momentum space correlation function on an infinite
 lattice
\begin{eqnarray}
 P(p) = \sum_{x,y,\nu}\langle M^\nu(x) M^\nu(y)\rangle
    \, e^{i p\cdot (x-y)}/N_l
\end{eqnarray}
 is shown in Fig.\ref{fig:chi} for $\beta$ values on either side of the
 percolation threshold and at the threshold itself.
 The zero momentum limit of this
 correlation function is the monopole susceptibility $\chi_s$. (Note
 that the limits have been taken in the correct order
 (volume goes to infinity before momentum goes to zero) needed
 to give a quantity that does not depend on the lattice boundary
 conditions.) The correlation doesn't diverge as the momentum goes
 to zero, in fact it vanishes quadratically at all values of the
 coupling. The susceptibility is identically zero for all $\beta$
 values. Zero susceptibility means that monopoles are
 perfectly shielded, which suggests confinement
 of {\it magnetic} charge at all values of the coupling, making
 the vacuum a dual insulator, not a dual superconductor.

  Many other quantities are also calculable.
 Particularly simple to calculate are the compact Wilson loops
 which again show no singularities, and give a potential between
 charges which is Coulombic at all $\beta$ values, inconsistent
 with confinement.
 The behaviour of the compact Wilson loops is particularly
 important in considering what happens if charged fermions are
 added to the theory, because the Wilson loops are the only
 feature of the gauge sector to which fermions can couple.
 If the Wilson loops are unaffected by the monopole percolation
 threshold then so are the fermions, and the fermion chiral
 phase transition can not be driven by the monopoles.

\section{Condensation and Percolation \label{perc_sect} }

   We have arrived at an apparent paradox. The paper \cite{newquench}
 looks very carefully at the Hands and Wensley monopole susceptibility
 \cite{Hands} making a careful finite size analysis on large lattices
 (up to $20^4$) and establishes that it shows every sign of diverging
 as lattice size goes to infinity.
 On the other hand we appear to have excellent evidence showing
 that the monopoles are in the same physical state on both sides
 of this apparent transition and that monopole properties vary
 analytically in the entire $\beta$ range.

   The resolution is simple - percolation has no connection with
 condensation or with any other field-theoretic or thermodynamic
 property of the theory.

  To see this consider the differences between the cluster
 susceptibility and other susceptibilities.
  Usually susceptibilities
 are the zero momentum limits of Greens functions, (so that it is clear
 that a divergent susceptibility implies a divergent correlation length
 and the existence of a continuum limit). On the other hand the cluster
 susceptibility is not given by a Greens function.
 This can be seen by looking at the definition \cite{Hands}
 \begin{eqnarray}
     \chi_c = \left<  \frac
  {\left(\sum_{n=4}^{n_{\mbox{\tiny{max}}}} g_n n^2\right)
                             - n_{\mbox{\small{max}}}^2 }
   {n_{\mbox{tot}}}  \right>
 \label{cluster}
 \end{eqnarray}
 (Here $n$ is the number of dual sites in a cluster linked together
 by monopole world lines, $g_n$ is the number of clusters of size $n$,
 $n_{\mbox{\small{max}}}$ is the size of the largest cluster and
 $n_{\mbox{\small{tot}}}$ is the total number of dual sites with
 monopole world lines passing through them.) Greens functions depend
 only on field values at their external legs, and so take no notice of
 clusters. Correlations that
 treat indistinguishable particles differently depending on cluster
 properties violate the interchange (Bose or Fermi) symmetries and so
 can not influence physical observables. (As an example of the
 unphysical nature of such a correlation consider an electron
 and positron. No measurement can distinguish an electron
 and positron that are on the same world line (because they were
 pair created together or because they will annihilate at some
 time in the distant future) from an electron and positron
 which are on different world lines.)

    Similarly measurements on a particle don't allow us to make
 any statements on the length of its world line, (old electrons
 and young electrons have exactly the same properties).
 However the percolation threshold can only have physical
 consequences if monopoles that are part of a large cluster
 have different interactions than those that are part of
 a small cluster.

 \section{Conclusion}

   In section \ref{analytic} we have seen that there is no phase
 transition, and in section \ref{perc_sect} seen that we should not
 have expected that percolation measurements would imply a phase
 transition

    In quenched, non-compact QED there is a percolation threshold
 at a $\beta \approx 0.244$ but this threshold does not signal
 monopole condensation or any other phase transition. If there
 is not a genuine phase transition at this point then there is
 no reason to think that monopoles are present in the continuum
 limit, and so no reason to discard lattice QED. Quenched QED
 provides a counter-example to the notion that a divergent
 cluster susceptibility implies a phase transition.

   What happens when fermions are added? Investigations of the
 percolation threshold \cite{nf} with two and four fermion flavours
 find that the percolation threshold looks very similar to the
 quenched case, occurring at almost the same monopole density
 and with almost identical critical exponents. This strong similarity
 suggests the same lack of physical implications. The general arguments
 on the lack of field-theoretic significance of percolation given in
 section \ref{perc_sect} do not depend on the theory being quenched,
 and so should should still apply.

     The papers on monopoles in non-compact QED
 \cite{newquench,Hands,confine,nf} do not prove that monopoles
 are relevant in the continuum limit of the lattice theory,
 and so do not invalidate the picture of the chiral phase
 transition presented in \cite{us}.


\begin{thebibliography}{monobib}
\bibitem{newquench} A. Koci\'c, J.B. Kogut and S.J. Hands,
   Phys. Lett. {\bf B289} (1992) 400.
\bibitem{Hands} S. Hands and R. Wensley,
        Phys. Rev. Lett. {\bf 63} (1989) 2169.
\bibitem{confine} S.J. Hands, J.B. Kogut, R. Renken, A. Koci\'c,
                  D.K. Sinclair and K.C. Wang,
      Phys. Lett. {\bf B261} (1991) 294.
\bibitem{us} M. G\"ockeler, R. Horsley, P. Rakow, G. Schierholz
             and R. Sommer,
         Nucl. Phys. {\bf B371} (1992) 713; \\
         P.E.L. Rakow,  Nucl. Phys. {\bf B 356} (1991) 27.
\bibitem{DGT}  T.A. DeGrand and D. Toussaint,
        Phys. Rev. {\bf D22} (1980) 2478.
\bibitem{percol} D. Stauffer, Phys. Rep. {\bf54} (1979) 1; \\
       G. Grimmett, Percolation (Springer-Verlag, 1989).
\bibitem{nf} S.J. Hands, R.L. Renken, A. Koci\'c, J.B. Kogut,
                  D.K. Sinclair and K.C. Wang,
               Illinois preprint Ill-(TH)-92-\#16 (1992); \\
         A. Koci\'c, J.B. Kogut and K.C. Wang,
               Illinois preprint Ill-(TH)-92-\#17 (1992).
\end{thebibliography}
\end{document}